\documentclass[trackchanges,twocolumn]{aastex7}
\usepackage[T1]{fontenc}
\usepackage{hyperref}
\newcommand{\unit}[1]{\ensuremath{\, \mathrm{#1}}}

\newcommand{\resP}{$P=331.52\pm0.02$ $\unit{\mathrm{days}}$} 

\newcommand{\resEcc}{$e=0.9462\pm0.0017$}

\newcommand{\resmsini}{$m_2$sin$i =  0.740\pm0.023$ $\unit{M_{\mathrm{Jup}}}$ }

\begin{document}

\title{The Closest Super-Eccentric Giant Planet: Dynamical History and Orbital Future of Gaia-3b}

\newcommand{\PSUAA}{Department of Astronomy \& Astrophysics, Pennsylvania State University, University Park, PA, 16802, USA}
\newcommand{\PSUCEHW}{Center for Exoplanets and Habitable Worlds, Pennsylvania State University, University Park, PA, 16802, USA}
\newcommand{\PSUICDS}{Institute for Computational and Data Sciences, The Pennsylvania State University, University Park, PA, 16802, USA}
\newcommand{\PSUStats}{Center for Astrostatistics, 525 Davey Laboratory, The Pennsylvania State University, University Park, PA, 16802, USA}
\newcommand{\UAm}{Anton Pannekoek Institute for Astronomy, 904 Science Park, University of Amsterdam, Amsterdam, 1098 XH}
\newcommand{\PSUARC}{Astrobiology Research Center, 525 Davey Laboratory, 251 Pollock Road, Penn State, University Park, PA, 16802, USA}
\newcommand{\PSETI}{Penn State Extraterrestrial Intelligence Center, 525 Davey Laboratory, 251 Pollock Road, Penn State, University Park, PA, 16802, USA}
\newcommand{\UA}{Steward Observatory, University of Arizona, 933 N.\ Cherry Ave, Tucson, AZ 85721, USA}
\newcommand{\UAA}{Department of Astronomy and Steward Observatory, University of Arizona, Tucson, AZ 85721, USA}
\newcommand{\Penn}{Department of Physics and Astronomy, University of Pennsylvania, 209 S 33rd St, Philadelphia, PA 19104, USA}
\newcommand{\Caltech}{Department of Astronomy, California Institute of Technology, 1200 E California Blvd, Pasadena, CA 91125, USA}
\newcommand{\STScI}{Space Telescope Science Institute, 3700 San Martin Dr, Baltimore, MD 21218, USA}
\newcommand{\JHU}{Department of Physics and Astronomy, Johns Hopkins University, 3400 N Charles St, Baltimore, MD 21218, USA}
\newcommand{\GoddardESAL}{Exoplanets and Stellar Astrophysics Laboratory, NASA Goddard Space Flight Center, Greenbelt, MD 20771, USA}
\newcommand{\GoddardISTD}{Instrument Systems and Technology Division, NASA Goddard Space Flight Center, Greenbelt, MD 20771, USA}
\newcommand{\GSFC}{NASA Goddard Space Flight Center, Greenbelt, MD 20771, USA}
\newcommand{\NOAO}{U.S. National Science Foundation National Optical-Infrared Astronomy Research Laboratory, 950 N.\ Cherry Ave., Tucson, AZ 85719, USA}
\newcommand{\UW}{Wisconsin address goes here}
\newcommand{\Macquarie}{School of Mathematical and Physical Sciences, Macquarie University, Balaclava Road, North Ryde, NSW 2109, Australia}
\newcommand{\MacquarieCentre}{Astrophysics and Space Technologies Research Centre, Macquarie University, Balaclava Road, North Ryde, NSW 2109, Australia}
\newcommand{\NIST}{National Institute of Standards \& Technology, 325 Broadway, Boulder, CO 80305, USA}
\newcommand{\CUBoulderEE}{Electrical, Computer \& Energy Engineering, 440 UCB, University of Colorado, Boulder, CO 80309, USA}
\newcommand{\CUBoulder}{Department of Physics, 390 UCB, University of Colorado, Boulder, CO 80309, USA}
\newcommand{\JPL}{Jet Propulsion Laboratory, California Institute of Technology, 4800 Oak Grove Drive, Pasadena, California 91109}
\newcommand{\MIT}{Kavli Institute for Astrophysics and Space Research, Massachusetts Institute of Technology, Cambridge, MA, USA}
\newcommand{\UCI}{Department of Physics \& Astronomy, The University of California, Irvine, Irvine, CA 92697, USA}
\newcommand{\Carleton}{Carleton College, One North College St., Northfield, MN 55057, USA}
\newcommand{\Carnegie}{Earth and Planets Laboratory, Carnegie Science, 5241 Broad Branch Road, NW, Washington, DC 20015, USA}
\newcommand{\PSUICS}{Institute for Computational and Data Sciences, Penn State, University Park, PA, 16802, USA}
\newcommand{\PSUCASt}{Center for Astrostatistics, 525 Davey Laboratory, 251 Pollock Road, Penn State, University Park, PA, 16802, USA}
\newcommand{\NESSF}{NASA Earth and Space Science Fellow}
\newcommand{\Princeton}{Department of Astrophysical Sciences, Princeton University, 4 Ivy Lane, Princeton, NJ 08540, USA}
\newcommand{\RUSSELL}{Henry Norris Russell Fellow}
\newcommand{\IAS}{Institute for Advance Study, 1 Einstein Drive, Princeton, NJ 08540, USA}
\newcommand{\Tsinghua}{Department of Astronomy, Tsinghua University, Beijing 100084, China}
\newcommand{\FlatironCCA}{Center for Computational Astrophysics, Flatiron Institute, 162 Fifth Avenue, New York, NY 10010, USA}
\newcommand{\ETH}{ETH Zurich, Institute for Particle Physics \& Astrophysics, Zurich, Switzerland}
\newcommand{\TIFR}{Department of Astronomy and Astrophysics, Tata Institute of Fundamental Research, Homi Bhabha Road, Colaba, Mumbai 400005, India}
\newcommand{\UCLA}{Department of Physics \& Astronomy, University of California Los Angeles, Los Angeles, CA 90095, USA}
\newcommand{\UMd}{Department of Physics \& Astronomy, 1023 University Drive, University of Minnesota Duluth, Duluth, MN 55812, USA}
\newcommand{\UIUC}{Department of Astronomy, University of Illinois at Urbana-Champaign, Urbana, IL 61801, USA}
\newcommand{\Amherst}{Department of Physics and Astronomy, Amherst College, 25 East Drive, Amherst, MA 01002, USA}
\newcommand{\UTAustin}{Department of Astronomy, The University of Texas at Austin, 2515 Speedway, Austin, TX 78712, USA}
\newcommand{\KavliInst}{Kavli Institute for Astrophysics and Space Research, Massachusetts Institute of Technology, Cambridge, MA 02139, USA}
\newcommand{\MITPhysics}{Department of Physics, Massachusetts Institute of Technology, Cambridge, MA 02139, USA}
\newcommand{\DTU}{DTU Space, Elektrovej, Building 328, DK-2800 Kgs. Lyngby, Denmark}

\correspondingauthor{Evan Fitzmaurice}
\email{exf5296@psu.edu}

\author[orcid=0000-0003-0199-9699]{Evan Fitzmaurice}
\affiliation{\PSUAA}
\affiliation{\PSUCEHW}
\email{exf5296@psu.edu}

\author[0000-0001-7409-5688]{Guðmundur Stefánsson}
\affil{Astrophysics \& Space Center, Schmidt Sciences, New York, NY 10011, USA}
\affiliation{\UAm}
\email{gstefansson@schmidtsciences.org}

\author[0000-0001-9596-7983]{Suvrath Mahadevan}
\affiliation{\PSUAA}
\affiliation{\PSUCEHW}
\email{suvrath@astro.psu.edu}

\author[0009-0006-0871-1618]{Haedam Im}
\affiliation{\MITPhysics}
\affiliation{\MIT}
\email{haedamim@mit.edu}

\author[0000-0003-3130-2282]{Sarah Millholland}
\affiliation{\MITPhysics}
\affiliation{\MIT}
\email{sarah.millholland@mit.edu}

\author[0000-0002-3853-7327]{Rachel B. Fernandes}
\affiliation{\PSUAA}
\affiliation{\PSUCEHW}
\email{rbf5378@psu.edu}

\author[0000-0002-1092-2995]{Anne Dattilo}
\affiliation{\PSUAA}
\affiliation{\PSUCEHW}
\affiliation{\PSUICDS}
\email{adattilo@psu.edu}

\author[0000-0003-1762-8235]{Simon H. Albrecht}
\affiliation{Stellar Astrophysics Centre, Department of Physics and Astronomy, Aarhus University, Ny Munkegade 120, 8000 Aarhus C, Denmark}
\email{albrecht@phys.au.dk}

\author[0000-0002-6096-1749]{Cullen H.\ Blake}
\affiliation{\Penn}
\email{chblake@sas.upenn.edu}

\author[0000-0002-0212-4563]{Olivia Curtis}
\affiliation{\PSUAA}
\affiliation{\PSETI}
\affiliation{Institute for Gravitation and the Cosmos, The Pennsylvania State University, University Park, PA 16802, USA}
\email{ocurtis@psu.edu}

\author[0000-0003-1439-2781]{Megan Delamer}
\affiliation{\PSUAA}
\affiliation{\PSUCEHW}
\email{mmd6393@psu.edu}

\author[0000-0002-3610-6953]{Jiayin Dong}
\affiliation{\FlatironCCA}
\affiliation{\UIUC}
\email{jdong@flatironinstitute.org}

\author[0000-0002-0078-5288]{Mark R.~Giovinazzi}
\affiliation{\Amherst}
\email{mgiovinazzi@amherst.edu}

\author[0000-0002-5463-9980]{Arvind F.\ Gupta}
\affiliation{\NOAO}
\email{arvind.gupta@noirlab.edu}

\author[0000-0003-1312-9391]{Samuel Halverson}
\affiliation{\JPL}
\email{samuel.halverson@jpl.nasa.gov}

\author[0000-0001-9825-7418]{Kamil K. Kalinowski}
\affiliation{Stellar Astrophysics Centre, Department of Physics and Astronomy, Aarhus University, Ny Munkegade 120, 8000 Aarhus C, Denmark}
\email{kamil@phys.au.dk}

\author[0000-0001-8401-4300]{Shubham Kanodia}
\affiliation{\Carnegie}
\email{skanodia@carnegiescience.edu}

\author[0000-0001-9626-0613]{Daniel M.\ Krolikowski}
\affiliation{\UA}
\email{krolikowski@arizona.edu}


\author[0000-0002-9632-9382]{Sarah E.\ Logsdon}
\affiliation{\NOAO}
\email{sarah.logsdon@noirlab.edu}

\author[0000-0003-2173-0689]{Marcus L. Marcussen}
\affiliation{Stellar Astrophysics Centre, Department of Physics and Astronomy, Aarhus University, Ny Munkegade 120, 8000 Aarhus C, Denmark}
\email{marcuslmarcussen@gmail.com}

\author[0000-0003-0241-8956]{Michael W.\ McElwain}
\affiliation{\GoddardESAL} 
\email{michael.w.mcelwain@nasa.gov}

\author[0000-0001-8720-5612]{Joe P.\ Ninan}
\affiliation{\TIFR}
\email{indiajoe@gmail.com}

\author[0000-0003-1324-0495]{Leonardo A.\ Paredes}
\affiliation{\UA}
\email{lparedes@arizona.edu}

\author[0000-0002-2488-7123]{Jayadev Rajagopal}
\affiliation{\NOAO}
\email{jayadev.rajagopal@noirlab.edu}

\author[0000-0003-0149-9678]{Paul Robertson}
\affiliation{\UCI}
\email{paul.robertson@uci.edu}

\author[0000-0001-8127-5775]{Arpita Roy}
\affiliation{Astrophysics \& Space Institute, Schmidt Sciences, New York, NY 10011, USA}
\email{arpita308@gmail.com}

\author[0000-0002-4046-987X]{Christian Schwab}
\affiliation{\Macquarie}
\email{mail.chris.schwab@gmail.com}

\author[0000-0002-4265-047X]{Joshua N. Winn}
\affiliation{Department of Astrophysical Sciences, Princeton University, Princeton, NJ 08544, USA}
\email{jnwinn@princeton.edu}

\author[0000-0001-6160-5888]{Jason T.\ Wright}
\affiliation{\PSUAA}
\affiliation{\PSUCEHW}
\affiliation{\PSETI}
\email{astrowright@gmail.com}

\begin{abstract}
With new high-resolution spectra and radial velocities (RVs) from the NEID spectrometer, we reaffirm the planetary designation of the previously identified super-eccentric giant planet Gaia~3b. Our additional 23 RVs span 356 days, including an RV observation obtained within hours of the periastron passage. By modeling all of the available RVs we collapse a previously tri-modal orbital period posterior distribution and report a more precisely measured value of \resP. Our measure of the eccentricity (\resEcc) is in agreement with previous measurements and thus corroborates the status of Gaia~3b as the third most eccentric planet known to date. We generate synthetic realizations of the Gaia astrometric time series and derived RUWE to show that only the planetary-companion and close to equal-mass binary cases are plausible when compared to the reported RUWE. We then show that the close to equal-mass binary case is unlikely based on the Gaia color-magnitude position and the shape and amplitude of the bisectors measured from the NEID cross-correlation functions. The confirmation of Gaia~3b's planetary status makes it the closest known super-eccentric giant planet ($d \approx 35$ pc) and leads us to anticipate that Gaia DR4 will detect more super-eccentric planets orbiting nearby Sun-like stars. Finally, due to the presence of a distant stellar companion, we demonstrate that a plausible pathway exists for excitation of the planet's eccentricity by von Zeipel-Kozai-Lidov oscillations.

\end{abstract}


\section{Introduction}\label{sec:intro}
Giant planets on super-eccentric orbits ($e > 0.9$) with small periapses are potential progenitors of hot Jupiters and provide a testbed for exploring the dynamical pathways that may lead them to short period orbits \citep{Socrates2012}. The detection of such planets has been limited, with only seven super-eccentric giant planets known to date (see Table \ref{tab:lit_planets} for a comparison of these known planets). As more super-eccentric giant planets are discovered, population level studies become viable as a means to understand the underlying exoplanet eccentricity distribution, as well as to decipher the conditions that lead to the observed population of hot Jupiters. 

The dynamical pathways that lead to the existence of hot Jupiters are reviewed in \citet{Dawson2018}. These dynamical pathways fall under two broad camps of migration: first, migration via interaction with the protoplanetary disk \citep{Goldreich1980}, and, second, high eccentricity migration caused by different dynamical interactions generating eccentric orbits, including von Zeipel-Kozai-Lidov oscillations \citep{vonZeipel1910,Kozai1962,Lidov1962} induced by a stellar mass \citep{Wu2003,FabryckyTremaine2007,Naoz2012} or planetary mass companion \citep{Lithwick2011,Naoz2011},  planet-planet scattering \citep{Rasio&Ford1996,Ford2006,Ford2008}, or secular chaos \citep{Wu2011}. The eccentric planetary orbits generated by these interactions are then expected to circularize due to tidal interactions between the planet and the host star.

The Gaia~3 (HIP 66074) system consists of a $0.705^{+0.025}_{-0.023}$ M$_{\odot}$, mid-K dwarf primary star, a widely separated ($44$\arcsec, 1550 au) L2-spectral type stellar companion that sits just above the hydrogen burning limit ($0.076\pm0.001$ M$_{\odot}$, $T_{\text{eff}}=2080\pm260$ K, \citet{Gomes2013}), and the super-eccentric giant planet candidate, Gaia~3b \citep{Sozzetti2023-Gaia3b}. A potential sub-stellar candidate was initially flagged by the non-single star/two-body astrometric solution released in Gaia Data Release 3 \citep[DR3,][]{GaiaDR3,gaia,Halbwachs2022,holl2022}. The single-Keplerian astrometric orbital solution suggested a dark (contributes no flux), sub-stellar companion on a $301\pm5$ day orbit with an eccentricity of $0.45\pm0.22$ and a mass $<$20 M$_\mathrm{Jup}$.

In a follow-up of Gaia exoplanet candidate two-body solutions, \citet{Winn2022} presented a radial velocity (RV)-only fit for Gaia~3b using a set of 10 RVs measured by the High Resolution Echelle Spectrograph \citep[HIRES,][]{Butler2017_HIRES}. They highlighted a factor of 15 (4.4$\sigma$) discrepancy in their fit RV semi-amplitude from the value reported in the Gaia two-body solution, making it unclear what the true minimum mass is. They were not able to identify the cause of the discrepancy since the raw astrometric measurements were not released in DR3. \citet{Sozzetti2023-Gaia3b} obtained more RVs with the HARPS-N spectrograph spanning just over the $\sim$300 day orbital period as predicted by the RV-only and Gaia-only orbital solutions. Their investigation revealed that the eccentricity of the sub-stellar companion was substantially higher ($e = 0.948 \pm 0.004$) than previously suggested by the Gaia solution ($e = 0.46 \pm 0.17$). Through numerical simulations of synthetically generated astrometric measurements, \citet{Sozzetti2023-Gaia3b} resolved the eccentricity discrepancy by showing that the orbital eccentricity is systematically underestimated due to the small astrometric perturbation and one dimensional nature of the Gaia along-scan measurements. They use the projected semi-major axis from the Gaia orbital solution to put a lower limit on the inclination, and thus an upper limit on the mass and conclude that the mass of Gaia~3b is between 3-7 M$_{\mathrm{Jup}}$. However, the Gaia astrometric orbital solution itself was later rescinded after a software bug was discovered but not detailed \footnote{https://www.cosmos.esa.int/web/gaia/dr3-known-issues}. The retraction of the solution makes it unclear at what confidence conclusions on the mass of Gaia~3b can be drawn from the reported values. 

In a NEID campaign following up Gaia astrometric exoplanet candidates \citep[see][]{Stefansson2025AJ_Gaia4_5b}, we observed Gaia~3b across an additional periastron passage to collapse the tri-modal orbital period posterior distribution in  \citet{Sozzetti2023-Gaia3b} that explains their reported 6.5\% uncertainty on the orbital peirod. We successfully obtained an RV measurement within hours of the periastron passage thanks to time-critical observations from the NEID guaranteed time observations program. With the addition of our NEID RVs we have constrained the true orbital period to be \resP, a now $\sim$0.006\% uncertainty. We report updated measures of the minimum mass, \resmsini, and eccentricity, \resEcc, that are consistent with the previous measurements in \citet{Sozzetti2023-Gaia3b}.

\citet{marcussen2023} followed up on Gaia astrometric exoplanet candidates, including Gaia~3b, using spectroscopy to attempt to rule out astrophysical false positives like spectroscopic binaries. From their analysis of the cross-correlation functions (CCFs) derived from the HIRES spectra they claimed to rule out the case of an equal mass binary, although in retrospect the spectra only covered epochs at very low velocity separation and were potentially not sensitive to noticeable changes in the CCFs. With the CCFs derived from our NEID spectra that fully sample the periastron passage, we revisit the possibility of a face-on, equal mass binary masquerading as a planetary companion and attempt to further rule out this scenario.

We describe the set of spectroscopic observations and subsequently derived RVs used in our analysis in Section \ref{sec:observations}, followed by the modeling and orbital parameter estimation for Gaia~3b in Section \ref{sec:orbital_solution}. In Section \ref{sec:astrom_plus_rvs}, we use the updated orbital solution to place limits on the mass of Gaia 3b based on astrometry from \textit{Hipparcos}, synthetically generated Gaia data, and our analysis of the shape changes in the NEID CCFs. We examine the implications of confirming Gaia 3b as a planet on our understanding and expectations of the underlying population of super-eccentric planets within 86 pc in Section \ref{sec:population}. With our updated orbital parameters, we re-evaluate the impact of the L2 stellar companion and the von-Zeipel-Kozai-Lidov mechanism on the observed eccentricity of Gaia~3b in Section \ref{sec:dynamical_analysis}. We discuss possible areas of interest for future follow-up and our current conclusions in Section \ref{sec:conclusions}.


\section{Spectroscopic Observations} \label{sec:observations}

\subsection{Archival Radial Velocities}\label{subsec:archival_rvs}
\subsubsection{HIRES}
We included 10 of the 11 publicly available HIRES RVs \citep{Butler2017_HIRES} (omitting, as in \citet{Winn2022}, the spectrum gathered at epoch JD~=~2455042.76395 due to the low signal-to-noise of the spectrum) in the modeling of the Keplerian signal. HIRES has a resolving power of $\sim60{,}000$ and a wavelength range from 370-800 nm. The HIRES RVs span nearly 1400 days from April 6, 2009 to February 3, 2013. The median RV uncertainty is 1.52 $\mathrm{m\,s^{-1}}$. 

\subsubsection{HARPS-N}
\citet{Sozzetti2023-Gaia3b} conducted an RV campaign of Gaia~3b with the HARPS-N spectrograph \citep{HARPSN_Consentino2012} that we summarize below. HARPS-N is a high-resolution (R $\sim$ $115{,}000$) optical spectrograph (380-690 nm) on the 3.58-m Telescopio Nazionale
Galileo in the Canary Islands (Spain). They collected 60 spectra between June 4th, 2022 and April 30th, 2023 (330 days). A typical visit consisted of a single 900 second exposure. Spectra were reduced with version 3.7.1 of the HARPS-N Data Reduction Software which is maintained by the Italian Centre for Astronomical Archives (IA2 \footnote{\url{https://www.ia2.inaf.it/}}). RVs were extracted with version 1.8 of the Template Enhanced Radial Velocity Reanalysis Application pipeline \citep{anglada-escude2012}. The median uncertainty of these RVs is  $\sigma_{\rm{RV}} = 1.25$ $\mathrm{m\,s^{-1}}$.

\subsection{NEID Radial Velocities}\label{sec:neid_rvs}
NEID is a fiber-fed, environmentally stabilized echelle spectrograph on the  WIYN 3.5 meter telescope at Kitt Peak National Observatory\footnote{The WIYN Observatory is a joint facility of the NSF’s National Optical Infrared Astronomy Research Laboratory, Indiana University, the University of Wisconsin–Madison, Pennsylvania State University and Princeton University.} \citep{Halverson2016,Schwab2016,Robertson2019}. NEID has wavelength coverage from 380 to 930 nm and resolving power $R \sim 110,000$. 

We acquired 23 RVs with NEID that span from November 23rd, 2023 to November 12th, 2024 ($356$ days). This interval spans 2 separate ``RV Eras'' where the instrument underwent a zero-point shift that manifests as an overall RV offset due to an upgrade of the laser frequency comb (LFC) photonic crystal fiber (PCF). The PCF upgrade extended the wavelength coverage of the LFC and expanded the number of echelle orders that use the LFC for wavelength calibration. More information can be found on the NEID Data Reduction Pipeline documentation site\footnote{https://neid.ipac.caltech.edu/docs/NEID-DRP/rveras.html}.
This means we have to treat data from different eras as effectively coming from separate instruments. 12 RVs prior to August 19, 2024 are considered NEID Run 2, while 11 RVs after August 24, 2024 are considered NEID Run 3. On November 11th and 12th, 2024, we acquired two visits per night, separated by $10$ hours to maximize coverage across the rapid periastron passage. 

Each visit consisted of one continuous 900 second exposure. The median signal-to-noise ratio per pixel is 42 measured at 550 nm. The spectral data are reduced using v1.4 of the NEID Data Reduction Pipeline. 
We extract RVs using a version of the template matching code \texttt{SERVAL} \citep{zechmeister2018} adapted for use with NEID spectra \citep{stefansson2022}. We performed separate SERVAL extractions for the Run 2 and 3 spectra. The median uncertainty of the SERVAL NEID RVs is $1.23$ $\mathrm{m\,s^{-1}}$. Our NEID RVs are included in Appendix \ref{appendix:RVs} in Table \ref{table:neid_rvs}. 

\begin{figure*}[]
    \centering    
    \includegraphics[width=\textwidth]{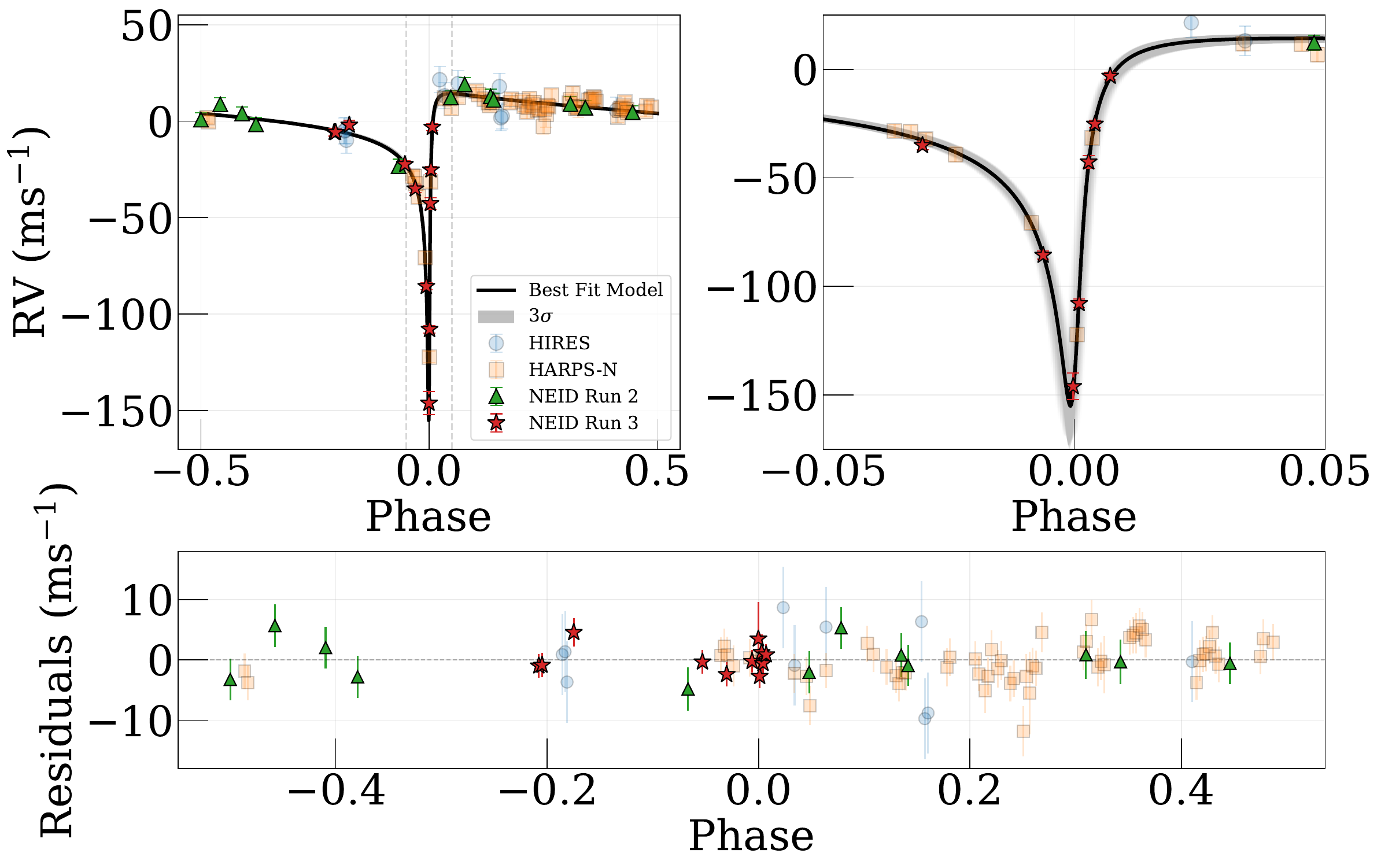}
    \caption{\textbf{Top Panels:} Best-fit RV model phase-folded and centered on 0 which corresponds to the time of periastron. The shaded region shows the 3$\sigma$ range of our fit. The vertical gray lines mark the phase space displayed in the right panel to better show our coverage of the periastron approach. \textbf{Bottom Panel:} Residuals after subtracting off the best-fit model from the data.}
    \label{fig:rv_plot}
\end{figure*}
\section{Orbital Solution} \label{sec:orbital_solution}

To determine the orbital solution, we used v6.0.0 of \texttt{Octofitter.jl}\footnote{https://github.com/sefffal/Octofitter.jl} \citep{Thompson2023_Octofitter} to sample the posterior distributions of our orbital parameters using stabilized variational non-reversible parallel tempering \citep{Surjanovic2022,Syed2021} implemented in Julia in \texttt{Pigeons.jl} \citep{Surjanovic2023}. This procedure was preferred over the Hamiltonian Monte Carlo sampler because of the computational efficiency and the ability to effectively sample high-dimensional, multi-modal posterior distributions. 

To build the Keplerian model we sample the orbital period, $P$, in years, $\sqrt{e}\textrm{cos}(\omega)$ and $\sqrt{e}\textrm{sin}(\omega)$, the minimum mass of the companion, $m_2$sin$i$, in Jupiter masses, the mass of the host star, $M_{*}$, in solar masses, and the epoch of periastron passage, $\tau$, expressed as a fraction of the orbital period that has passed since a specified reference date \citep{Blunt2020_orbitize}, which in our case was $t_{\rm{ref}} = 2459950.5$. $\tau$ is then $(t_{\rm{peri}}-t_{\rm{ref}})/P$. The semi-major axis, $a$, in au, the eccentricity, $e$, the argument of periastron, $\omega$, in radians, the minimum mass of the companion, $m_2\textrm{sin}i$, and the time of periastron, $t_{\textrm{peri}}$ are derived from the sampled parameters.


For each RV instrument we add parameters for an RV offset, $\gamma_{\textrm{inst}}$, and a white-noise jitter term, $\sigma_{\textrm{inst}}$, both in units of $\mathrm{m\,s^{-1}}$. As explained in Section \ref{sec:neid_rvs}, we treat data from NEID Runs 2 and 3 as individual instruments. As shown in Table \ref{tab:priors_and_posteriors}, we set uniform priors on the majority of the sampled parameters, except for the primary star mass where we set a normal prior with a mean and standard deviation according to the constraint from \citet{Sozzetti2023-Gaia3b} of $0.705^{+0.025}_{-0.023}$ M$_{\odot}$ derived using YY-isochrones \citep{Yi2001}. The returned value for $M_*$ is just a reflection of the prior and is not constrained further by our data.

\citet{Sozzetti2023-Gaia3b} added a gaussian process (GP) using a quasi-periodic (QP) kernel in their fit to describe correlated noise induced by stellar activity. This was motivated by a prominent $\sim$35-day periodicity in the periodograms of the stellar activity metric, $S_{\mathrm{HK}}$, from the HARPS-N spectra, and TESS lightcurves, which are interpreted as the stellar rotation period. We tested the inclusion of a QP-GP in our fits but found that the median value of the posteriors both with and without the QP-GP were consistent while the uncertainties were reduced by less than 10\% when including the QP-GP. As a simpler and more conservative model, we report the results of the single Keplerian fit without the GP in Table \ref{tab:priors_and_posteriors}. We show our fit in Figure \ref{fig:rv_plot}. We also tested the inclusion of linear and quadratic terms, both together and separately, to model a long-term trend in the RVs that might be induced by the longer period L2-spectral type stellar companion, but the model excluding these terms was preferred as assessed by the Bayesian Evidence where $\ln (Z_{\textrm{no trend}}/Z_{\textrm{linear}}) \approx 1.5 $ and $\ln(Z_{\textrm{no trend}}/Z_{\textrm{linear+quad}}) \approx 4.4$.

We follow the guidance in the \texttt{Octofitter.jl} documentation to assess the convergence of the parallel tempering algorithm \footnote{https://sefffal.github.io/Octofitter.jl/dev/samplers/}.
We sample the posteriors using 14 rounds such that the Bayesian Evidence factor converged to a consistent value. We set the number of chains and variational chains both to 22 to be roughly twice the global communication barrier, $\Lambda$, as this helps ensure that sufficient swapping between the chains of different temperatures is achieved \citep{Syed2021}. We also visually inspected the progression of the swaps over the sampling progression to ensure that adequate traversing across the annealing space is achieved. The posteriors are drawn from the untempered chain. We compute convergence metrics on this chain such as the classic Gelman-Rubin statistic and the newer split-$\hat{R}$ \citep{Vehtari2021-Rhat} being within 1\% of unity for all parameters. Additionally, the effective sample size is much greater than 50 for all parameters. We finally confirm that the log-posterior density has converged onto a single target distribution as opposed to showing excursions that suggest a lack of convergence.

\begin{deluxetable}{lll}[t]  
\caption{Priors and Posterior Values for Gaia~3b\label{tab:priors_and_posteriors}}
\tabletypesize{\scriptsize}
\setlength{\tabcolsep}{3pt}
\renewcommand{\arraystretch}{0.95}
\tablewidth{0.48\columnwidth}   

\tablehead{\colhead{Parameter} & \colhead{Prior} & \colhead{Value}}

\startdata
\cutinhead{Sampled Orbital Parameters}
$P$ [days] & $\mathcal{U}(0,500)$ & $331.526 \pm 0.015$ \\
$\sqrt{e}\cos\omega$ & $\mathcal{U}(-1,1)$ & $0.854 \pm 0.011$ \\
$\sqrt{e}\sin\omega$ & $\mathcal{U}(-1,1)$ & $0.465 \pm 0.018$ \\
$M_\star$ [M$_{\odot}$] & $\mathcal{N}(0.705,0.025)$ & $0.706 \pm 0.025$ \\
$m_2\sin i$ [M$_{\mathrm{Jup}}$] & $\mathcal{U}(0.003,1000)$ & $0.740 \pm 0.024$ \\
$\tau$ & $\mathcal{U}(0,1)$ & $0.03650 \pm 0.00010$ \\
\cutinhead{Derived Orbital Parameters}
$t_{\rm peri}$ [days] & \nodata & $2459962.586 \pm 0.044$ \\
$e$ & \nodata & $0.9462 \pm 0.0017$ \\
$^{*}\omega_p$ [deg] & \nodata & $28.5 \pm 1.3$ \\
$K$ [$\mathrm{m\,s^{-1}}$] & \nodata & $84.7 \pm 2.7$ \\
\cutinhead{Instrumental Parameters}
$\sigma_{\rm NEID\,Run\ 2}$ [$\mathrm{m\,s^{-1}}$] & $\mathcal{U}(0.001,100)$ & $3.5 \pm 1.0$ \\
$\sigma_{\rm NEID\,Run\ 3}$ [$\mathrm{m\,s^{-1}}$] & $\mathcal{U}(0.001,100)$ & $1.7 \pm 1.2$ \\
$\sigma_{\rm HARPS-N}$ [$\mathrm{m\,s^{-1}}$] & $\mathcal{U}(0.001,20)$ & $2.77 \pm 0.36$ \\
$\sigma_{\rm HIRES}$ [$\mathrm{m\,s^{-1}}$] & $\mathcal{U}(0.001,20)$ & $6.9 \pm 2.1$ \\
$\gamma_{\rm NEID\,Run\ 2}$ [$\mathrm{m\,s^{-1}}$] & $\mathcal{U}(-100,100)$ & $-5.9 \pm 1.1$ \\
$\gamma_{\rm NEID\,Run\ 3}$ [$\mathrm{m\,s^{-1}}$] & $\mathcal{U}(-100,100)$ & $33.5 \pm 1.0$ \\
$\gamma_{\rm HARPS-N}$ [$\mathrm{m\,s^{-1}}$] & $\mathcal{U}(-100,100)$ & $-7.94 \pm 0.43$ \\
$\gamma_{\rm HIRES}$ [$\mathrm{m\,s^{-1}}$] & $\mathcal{U}(-100,100)$ & $-5.6 \pm 2.4$ \\ 
\enddata
\tablecomments{$^{*}$Figure $\ref{fig:rv_plot}$ uses $\omega_{\star}$, which is $180^{\circ}$ offset from $\omega_{p}$. This means in the figure that positive velocities are stellar red-shifts and negative velocities are stellar blue-shifts.}
\end{deluxetable}

\section{Mass Constraints from Astrometry} \label{sec:astrom_plus_rvs}
\subsection{An Upper Mass Limit from Hipparcos}
We attempt to further rule out the possibility that Gaia~3 is a face-on stellar binary by using the available astrometry to impose limits on the mass of Gaia~3b. We follow the process outlined in \citet{Kane2016} which placed an upper limit on the mass of the super-eccentric giant planet HD~20782b using the \textit{Hipparcos} intermediate astrometry data \citep{vanLeeuwen2007a,vanLeeuwen2007b}. The median single-measurement precision for HIP~66074/Gaia 3 is 7.6 mas, which is well above the minimum projected semi-major axis of Gaia~3b of 0.024 mas. 
Simulations to estimate detection sensitivity for Hipparcos in \citet{Sahlmann2011} showed that if the semi-major axis of an orbit is at least 70\% of the \textit{Hipparcos} measurement precision, the orbit is always detected at the 3$\sigma$ level. With the 7.6 mas measurement precision for Gaia-3, orbits above 5.3 mas would have been detectable at this level by \textit{Hipparcos}. Given this lower limit on the projected semi-major axis and our updated orbital parameter estimates, we can solve for inclination using Equation 9 of \citet{Sahlmann2011} and put an upper limit on the mass of Gaia~3b. We find an inclination lower limit of 0.25 degrees and subsequent upper mass limit of 199 Jupiter masses or 0.190 solar masses.  Using \textit{Hipparcos} data alone we are unable to rule out the possibility that Gaia~3b could be a brown dwarf or low mass star.

\subsection{Generative Modeling to Replicate the Gaia RUWE}
Given the withdrawal of the Gaia two-body astrometric orbital solution, we investigate what limits we can place on the secondary mass by synthetically generating the astrometric time series and calculating the renormalized unit weight error (RUWE) for each orbital configuration. The RUWE is a quality of fit metric for the astrometric fit to the along-scan observations. We compare the synthetic value to the reported RUWE from the single star only astrometric fit since that value was not retracted. For the simulations we use the \texttt{predict\_astrometry\_luminous\_binary} and \texttt{check\_ruwe} functions in \texttt{gaiamock}\footnote{https://github.com/kareemelbadry/gaiamock} \citep{gaiamock-El-Badry2024} which builds off of the astrometric uncertainty relationship in \citet{Holl2023b}. 

We draw 5000 samples directly from the posteriors of our updated RV fit for $P$, $e$, $\omega$, $M_{*}$, and the time of periastron. For the longitude of the ascending node ($\Omega$) we draw from a uniform distribution from [0, 2$\pi$]. For the astrometric parameters (right ascension, declination, proper motions, and parallax) and $G$-band magnitude, we use the reported values from Gaia DR3. The remaining input parameters are the inclination, mass of the secondary, and the flux ratio. We iterate over a range of 1000 mass ratios ($q$) evenly spaced from 0.002 to 1.0 and compute the RUWE distribution at each point using the 5000 samples. Given $q$ and the primary mass samples, the secondary mass is defined at each step. For $q < 0.1$, we assume the companion contributes no flux and use the corresponding $m_2\sin i$ posterior sample from the RV fit to determine the inclination. For $q >= 0.1$, the companion is no longer dark and the measured RVs could be flawed since we weren’t accounting for the secondary set of lines present in the spectra. To account for this we use a two gaussian model to simulate the cross-correlation functions from a spectroscopic binary and measure the new RV. The gaussians are scaled in intensity by the flux ratio and offset in velocity space based on the mass ratio and the line-of-sight projection due to the inclination. For a set of 10 mass ratios evenly spaced between 0.1 and 1.0, we use the MCMC package \texttt{emcee} \citep{emcee_ForemanMackey2013} to estimate the posterior distribution of the inclinations that could have reproduced the observed RVs given our two gaussian model. We then use an unweighted univariate spline fit to estimate a median and 1-sigma relationship between $q$ and the inclination to randomly draw from in our RUWE simulations. For mass ratios below 0.1, we assume the flux from the companion is negligible and set the flux ratio to $10^{-8}$. Above this level we use the relationship from \citet{Giovinazzi2022}\footnote{https://github.com/markgiovinazzi/gorp\_masses} to convert from mass to the absolute $G_{\text{RP}}$-band magnitude and then compute the flux ratio for that specific secondary mass. 

\begin{figure}[h]
    \centering    \includegraphics[width=1.0\columnwidth]{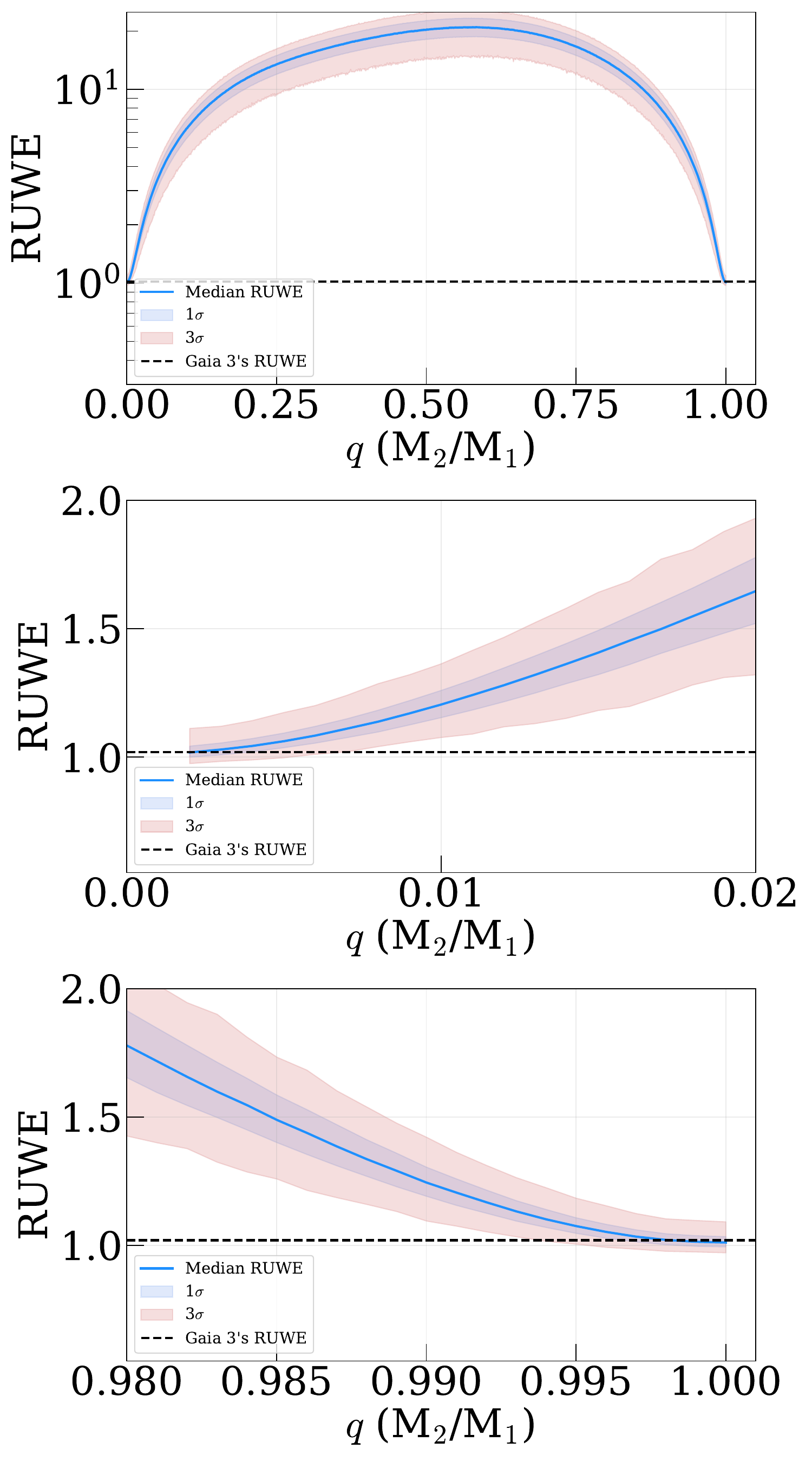}
    \caption{\textbf{Top:} Median, 1$\sigma$,and $3\sigma$ RUWE values for simulated astrometric time series as a function of the mass ratio, $q$. \textbf{Middle:} Zoomed in portion of the top panel showing plausible parameter space for a giant planet. \textbf{Bottom:} Zoomed in portion of the top panel showing plausible parameter space for an equal mass binary. The horizontal dashed black lines show Gaia~3's reported RUWE of 1.02. }
    \label{fig:ruwe_plot}
\end{figure}

Given the reported RUWE for Gaia~3 of 1.02, our simulations in Figure \ref{fig:ruwe_plot} show that there are only two plausible regimes of $q$ that could have produced a RUWE value consistent within 3$\sigma$. These are where $q < 0.007$ that results in a planetary mass companion between the minimum mass and up to $\approx$ 5 M$_{\mathrm{Jup}}$, and where $0.993 < q < 1.0$ that results in the case of a nearly face-on, close to equal mass binary.

\begin{figure}[h!]
    \centering
    \includegraphics[width=0.48\textwidth]{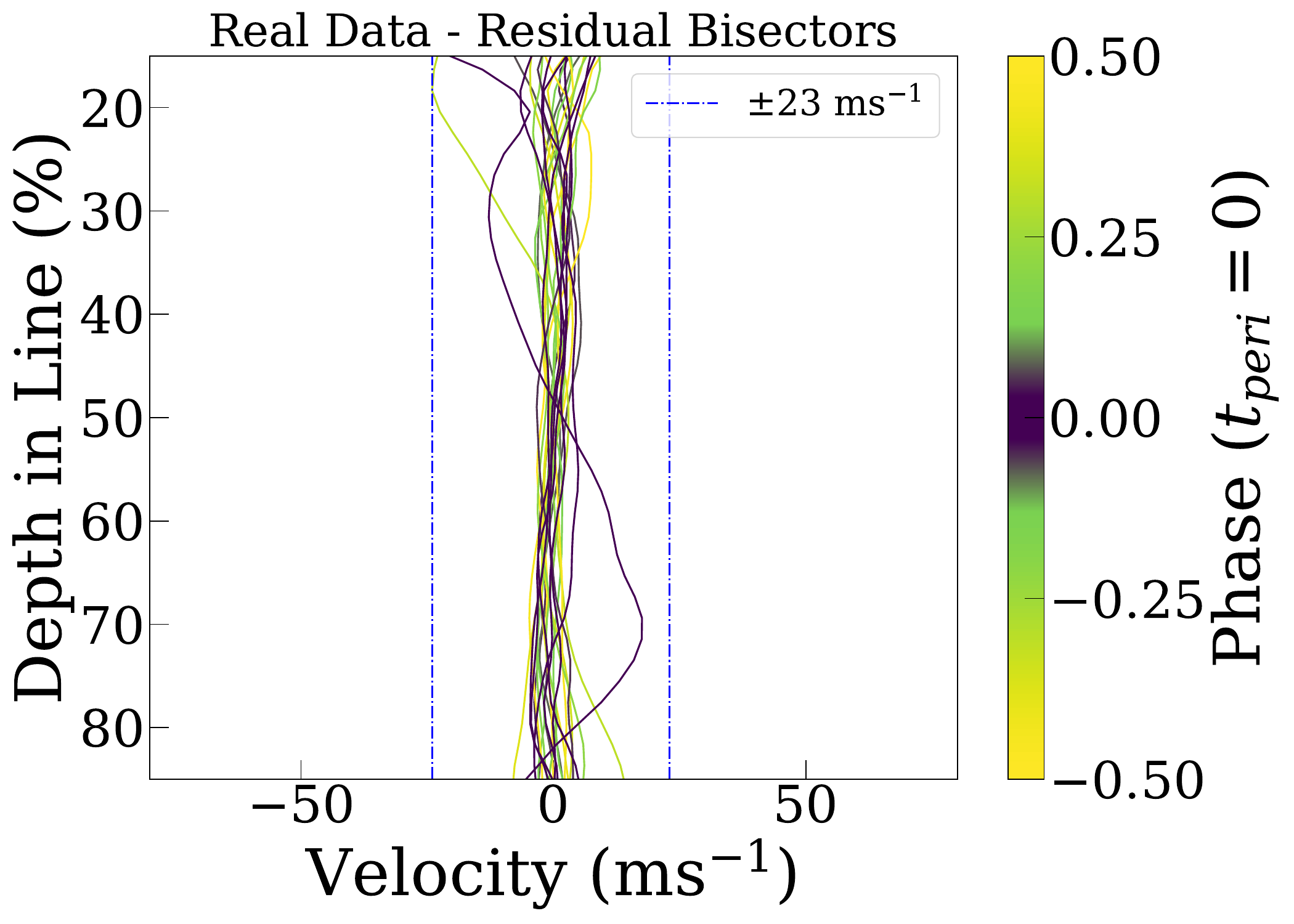}\\[1ex]
    \includegraphics[width=0.48\textwidth]{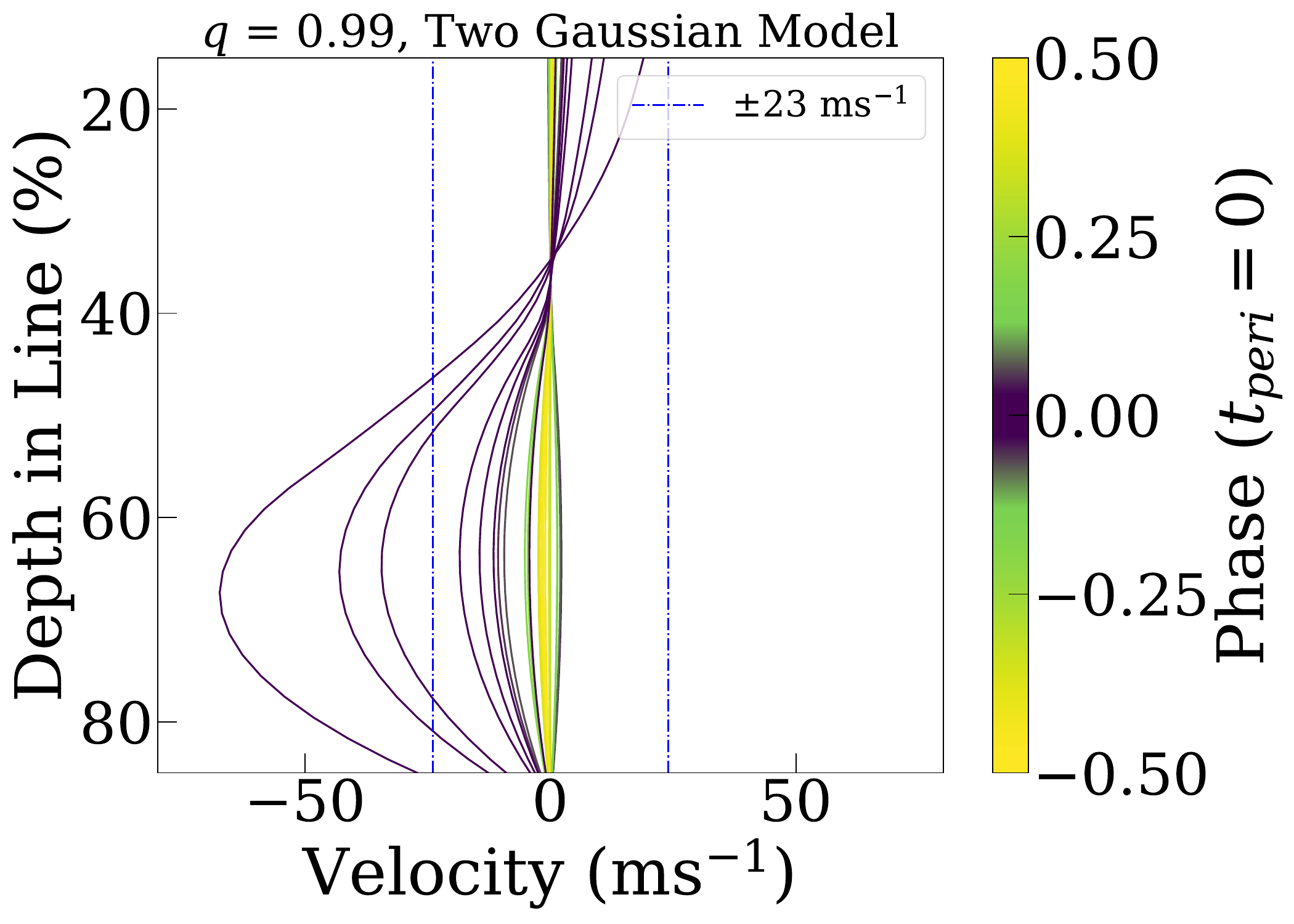}\\[1ex]
    \includegraphics[width=0.48\textwidth]{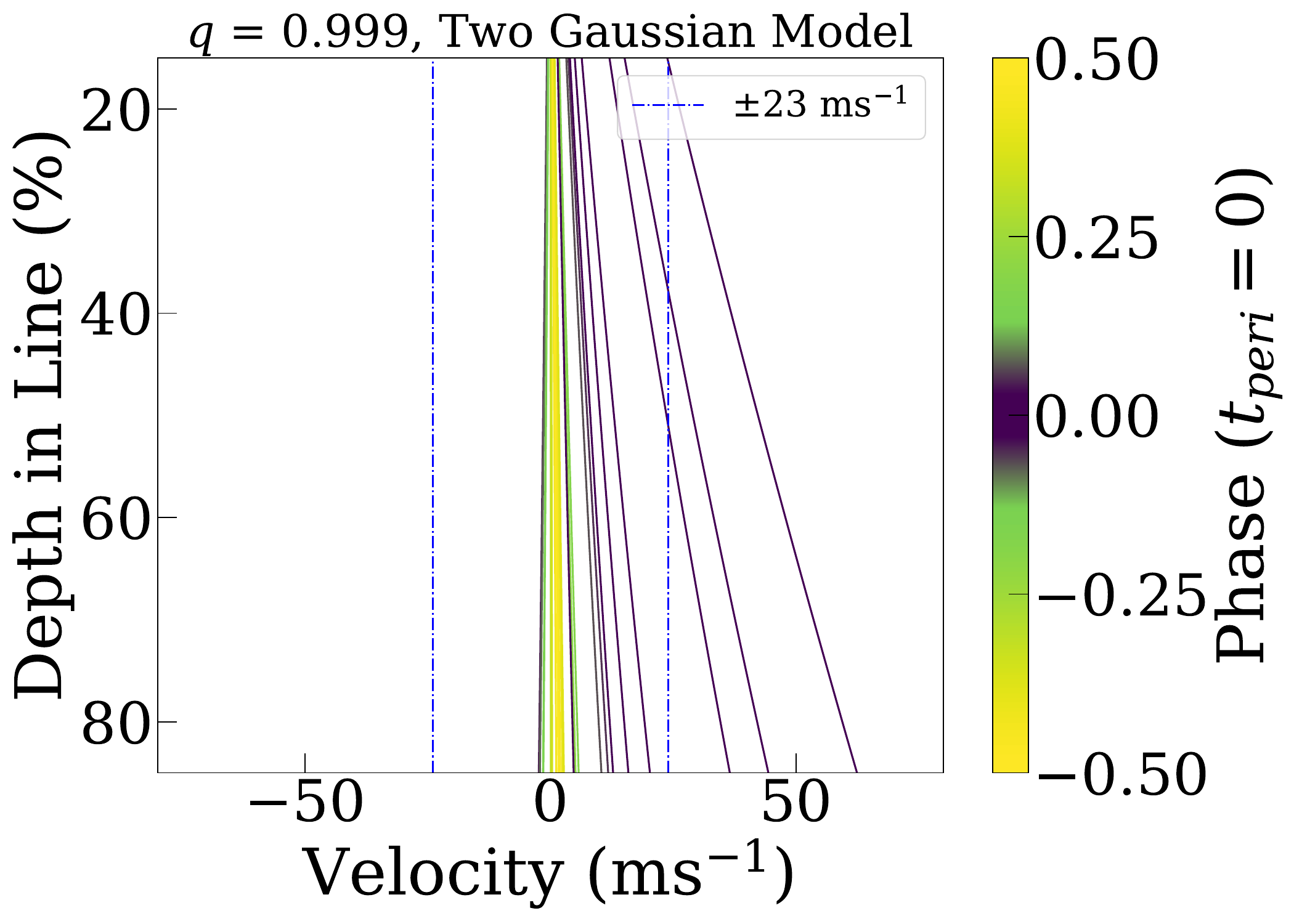}
    \caption{Measured bisectors from the actual NEID cross-correlation functions (\textbf{Top}) and our two gaussian model simulated cross-correlation functions at different mass ratios of $q = 0.99$ (\textbf{Middle}), $q = 0.999$ (\textbf{Bottom}). An average of the bisectors where the RV slope is low is subtracted to remove the characteristic bisector C-shape due to granulation. 
    }
    \label{fig:bisectors_plot}
\end{figure}
\begin{figure}[h!]
    \centering    \includegraphics[width=1.0\columnwidth]{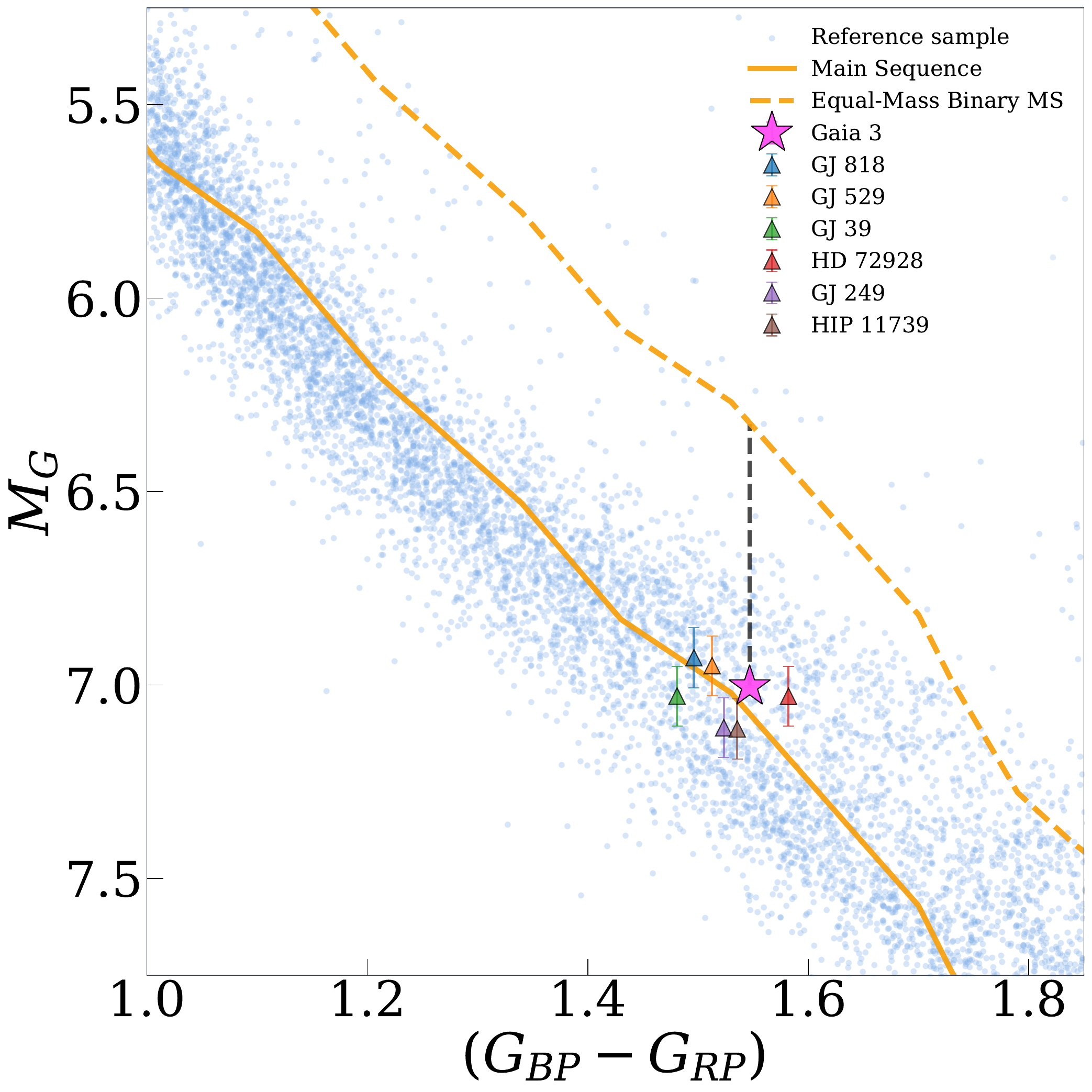}
    \caption{Based on this Gaia color-magnitude diagram along the main sequence (MS), Gaia 3 (pink star), with its reported parallax, does not present as an obvious over-luminous binary. The reference sample are Gaia DR3 targets within 100 parsecs and with a cut of RUWE<1.2. The colors and magnitudes for the MS and equal-mass binary MS lines come from \citet{Pecaut&Mamajek2013}. The vertical black dashed line connects to the position where we would expect Gaia 3 if it were an equal-mass binary. The other highlighted points are single standard stars of the same spectral type as Gaia 3 (K6V).}
    \label{fig:gaia_cmd}
\end{figure}

As noted previously, in the case of an equal mass binary the observed spectrum would be the combination of both spectra resulting in notable changes in the shapes of the CCFs extracted from our high-resolution spectra. Because the maximum velocity separation for this system ($\sim 300$ m\,s$^{-1}$) is much lower than the velocity resolution of NEID, we do not expect to see a double peak in the CCF which is characteristic of spectroscopic binaries. However, the shape changes of the CCF bisectors could reveal the nature of the nearby companion. We use the aforementioned two gaussian simulations of the CCFs to simulate the bisector measurements at our timestamps. The bisectors measured from the actual NEID spectra are compared to our simulations in Figure \ref{fig:bisectors_plot}. We showcase two $q$ values, 0.99 and 0.999, as these span the rejected and permitted region of $q$ space from our RUWE simulations (bottom panel Fig. \ref{fig:ruwe_plot}) The maximum absolute velocity swing we observe in the real bisectors is $\approx$23 m\,s$^{-1}$ while our simulations exceed this at both $q$ values. As we approach very close to $q$ of 1 the changes in the bisectors are expected to vanish as the two gaussian model becomes almost completely symmetric. Our model shows this expected behavior, with bisectors that do fall below the observed span of the real bisectors. The bisectors alone cannot formally rule out mass ratios within $<0.1\%$ of $q=1$. However, if Gaia-3 really were two nearly identical stars, we would expect it to show a significant over-luminosity on the HR diagram. In Figure \ref{fig:gaia_cmd}, we show Gaia~3's position on the Gaia color-magnitude diagram along with the positions of other single standard stars\footnote{\url{https://www.pas.rochester.edu/~emamajek/spt/K6V.txt}} of the same spectral type (K6V) to show that it is not anomalous by comparison. We plot the main sequence based on colors and magnitudes from \citet{Pecaut&Mamajek2013}. We also plot the expected position of the main sequence if the stars were actually equal mass binaries. Gaia~3's expected position if it were an equal-mass binary is connected by the vertical dashed black line. The uncertainties on the single standard stars are the sample standard deviation of the 6 stars. The reference sample is built on Gaia DR3 sources within 100 parsecs and with RUWE values $<1.2$.  

The upper mass limit from \textit{Hipparcos} and the combination of the Gaia color-magnitude position with the RUWE/CCF bisector analysis leads us to conclude that the case of an equal mass binary is unlikely and that we are in agreement with the planetary mass designation of Gaia~3b by \citet{Sozzetti2023-Gaia3b}.

However, we cannot formally rule out the possibility that the nearby companion has a high $v$sin$i$. In this case, the rotational broadening would wash out any contribution to the observed spectrum/CCFs and make it undetectable similar to the planet case. We believe this scenario is unlikely because at the age of the system, $t =7.9^{+4.9}_{-4.1}$ Gyr \citep{Sozzetti2023-Gaia3b}, a $\sim0.7 M_\odot$ mid-K dwarf would have likely spun down to a rotational velocity that would not prohibit its detection via our CCF inspection, though we acknowledge exceptions to the assumed relationship between age and rotational velocity \citep{Agueros2018,Curtis2019,Curtis2020}. In the case of a white dwarf companion where the flux ratio remains consistently small while the mass ratio increases, the RUWE would not turnover at higher $q$ and therefore would not have produced a RUWE value close to 1.

\section{The Underlying Population of Super-Eccentric Planets} \label{sec:population}
As the number of known super-eccentric ($e$ > 0.9) giant planets continues to grow it becomes possible to place order of magnitude constraints on the extreme end of the exoplanet eccentricity distribution and the true underlying population of such planets. Understanding the population is critical for evaluating the efficiency of high-eccentricity migration for the formation of hot Jupiters. The confirmation of Gaia~3b as a planet means there are now two super-eccentric giant planets (HD 20782b, $m_p\sin i =$ $1.4878^{+0.1045}_{-0.1066}$ M$_{\mathrm{Jup}}$, \citealt{Udry2019}) known within $\sim$36 pc of the Sun, and five total within $\sim$100 pc (Table \ref{tab:lit_planets}).

To understand how the presence of Gaia 3b impacts our expectations of the broader population, we can compare the rate of super eccentric giant planets needed to reproduce the sample when we include and exclude Gaia~3b. These calculations are intended as a first-order estimate, not formal occurrence-rate measurements as they neglect  completeness calculations that are highly technique and survey dependent. In reality, both RV and transit surveys are likely highly incomplete for super-eccentric planets due to the inhomogeneous temporal coverage and long period of these systems.

Based on Gaia color-magnitude limits there are $52{,}066$ FGK stars within 100~pc \citep{Gaia2021_GCNS, Pecaut&Mamajek2013}. \citet{Fernandes2019} report an occurrence rate of 26.6$^{+7.5}_{-5.4}$\% for planets with masses of $0.1$–$20\,M_\mathrm{Jup}$ and semi-major axes of $0.1$–$100$ au orbiting FGK stars. This implies $\sim13{,}849$  giant planets within 100~pc across all eccentricities. Assuming 100\% detection completeness for the super-eccentric planets already known within 100-pc while excluding Gaia~3b, the inferred fraction of giant planets on super-eccentric orbits is therefore $\sim$0.03\%. Within 36~pc, there are 2,432 FGK stars, corresponding to $\sim$647 giant planets. Applying the $0.03$\% super-eccentric fraction yields an expectation of $0.2 \pm 0.4$ planets. The presence of one super-eccentric giant planet within 36 pc, let alone two including Gaia~3b, exceeds this estimate. However, $0.2\pm0.4$ predicted planets is a lower limit, since we likely do not have 100\% detection completeness out to 100 pc. 

Alternatively, if we include Gaia~3b and assume that the super-eccentric planet sample is complete only within 36~pc, the implied super-eccentric fraction rises to $\sim$0.3\%. Extrapolating this rate to 100~pc predicts $43 \pm 7$ giant planets on super-eccentric orbits. While Gaia 3b is expected under this assumption, this value exceeds the 5 known super-eccentric giant planets within 100 pc.

This implies that more super-eccentric planets remain to be found. Given our expectation of the lack of completeness of RV and transit surveys, this is not entirely surprising. \textit{TESS} \citep{TESS_Ricker2015} lacks the observational baseline in most areas of the sky, while RV surveys are typically too expensive/not efficient enough given observing cadence and observational cost to have good enough completeness to catch many super-eccentric planets. Previous works have shown that the observed number of super-eccentric planets in the \textit{Kepler} sample is less than predicted \citep{Jackson2023,Dawson2012,Socrates2012}, but the \textit{Kepler} field may not be a suitable representation for the full volume of nearby Sun-like stars. The baseline, anticipated precision, and all-sky nature of Gaia DR4 is expected to provide a powerful opportunity in the near term to probe the true population of super-eccentric planets and to evaluate the dominant migration mechanisms in the formation of hot Jupiters \citep{Lammers2026}. High-resolution spectroscopy remains a key part of the puzzle in identifying false-positives \citep{Stefansson2025AJ_Gaia4_5b, marcussen2023,Winn2022} and filling out the rest of the orbital solution for these planets.

\section{Dynamical Analysis} \label{sec:dynamical_analysis}
The presence of a distant binary companion in the Gaia-3 system suggests that Gaia~3b could have reached its current close-in, super-eccentric orbit via high-eccentricity migration induced by von Zeipel-Lidov-Kozai (ZLK) oscillations. In this scenario, the inclined distant perturber excites oscillations in which the planet's orbital eccentricity and the mutual inclination (between the planet orbit and binary orbit) undergo periodic variations. Over successive ZLK cycles, tidal dissipation at periastron passages gradually damps the eccentricity, ultimately circularizing the planet's orbit via strong planet-star interactions \citep{Wu2003, Fabrycky2007, Dong2014, Anderson2016,
Dawson2018, Liveoak2026}.

Here, we present a plausible dynamical history of Gaia~3b using numerical simulations of high-eccentricity migration via ZLK oscillations under the assumption that Gaia~3b started on a wider orbit initially. 
These simulations are intended as a proof of concept that such a mechanism could operate in this system, given its known parameters and binary companion mass.  It is not as an exhaustive exploration of all possible pathways and parameter regimes. We use the open-source code \texttt{SecularMultiple}~\citep{HamersPortegiesZwart2016, Hamers2018, Hamers2020}, which integrates orbit-averaged equations of motion for hierarchical systems composed of nested binary orbits. \texttt{SecularMultiple} incorporates general relativistic (GR) precession and tidal evolution based on the constant time-lag equilibrium tide model \citep{EggletonKiseleva1998}. We adopt the octupole-order approximation for our dynamical simulation.


Before diving into simulations, we must identify a suitable set of initial conditions, which we do using methods similar to those described in \cite{Gupta2024}. We first note that there is a critical semi-major axis $a_{\mathrm{crit}}$ such that ZLK oscillations are quenched when the timescale of short-range forces (dominated by GR precession) of the planet's orbit is faster than that of the ZLK oscillations \citep{Liu2015}. We calculate this as \citep{Naoz2016}

\begin{equation}
a_{\mathrm{crit}} = \left[\frac{3 a_{2}^3 \left(1-{e_2}^2\right)^{3 / 2}G\left(M_{\star}+m_2\right)^{2}}{\left(1-e^2\right) M_3 c^2}\right]^{1/4},
\end{equation}
where $m_2$ and $e$ are the planetary mass and eccentricity of Gaia 3b, and $a_2$, $e_2$, and $M_3$ are the semi-major axis, eccentricity, and the mass of the binary companion, respectively. Using the initial eccentricities adopted in our simulations below (${e_{\mathrm{init}}} = 0.1$ for Gaia~3b and ${e_{2,\mathrm{init}}} = 0.1$ for the binary companion), we determine that the initial semi-major axis has to satisfy $a_{\mathrm{init}} \gtrsim 5.18$ au, otherwise ZLK oscillations would have been quenched due to the dominance of the short-range forces. This value is consistent with the findings of \citet{Sozzetti2023-Gaia3b}, who used analytical and numerical integrations to conclude that relativistic precession dominates over ZLK oscillations given Gaia~3b's current observed semi-major axis and statistical constraints on the eccentricity of the stellar binary companion. Thus, for this proof of concept, we adopt an initial semi-major axis $a_{\mathrm{init}} = 12$ au moving forward, though we note that other values are possible.

With this initial semi-major axis, we can determine the maximum eccentricity required for the initial cycle of ZLK-driven migration in order for the planet to reach small enough periapse distances and approach its present-day configuration. We use the conservation of $a_f \equiv a(1-e_{\max}^2)$, where $a_f$ is the final semi-major axis after complete tidal circularization ($e=0$) and $e_{\max}$ is the maximum eccentricity achieved during ZLK cycles. The required maximum eccentricity of the initial ZLK cycle $e_{\mathrm{init}, \max}$ is then given by
\begin{equation}
\label{eq: emax}
e_{\mathrm{init}, \max} = \sqrt{1-\frac{a_{\mathrm{pres}}(1-e_{\mathrm{pres}}^2)}{a_{\mathrm{init}}}},
\end{equation}
where $a_{\mathrm{pres}}$ and $e_{\mathrm{pres}}$ are the present-day semi-major axis and eccentricity of the planet. Using the resulting  $e_{\mathrm{init}, \max}$, we calculate the minimum initial inclination between the orbit of the planet and perturber to reach the required $e_{\mathrm{init}, \max}$ using Equation~(50) from \cite{Liu2015}
\begin{equation}
\label{eq: I_min}
\cos^2 I_{\mathrm{init}} = \frac{3j_{\mathrm{init},\min}^2}{5}\left[1 -  \frac{8\varepsilon_{\mathrm{GR}}}{9e_{\mathrm{init},\max}^2}\left(\frac{1}{j_{\mathrm{init}, \min}}-1\right)\right],
\end{equation}
where $j_{\mathrm{init}, \min} = \sqrt{1-e_{\mathrm{init}, \max}^2}$.
Here, we only consider GR precession, assuming it dominates over the contributions from tidal precession and rotational oblateness. $\varepsilon_{\mathrm{GR}}$ quantifies the contribution of GR precession relative to the quadrupole potential, defined by

\begin{figure}[h]
    \centering    \includegraphics[width=1.0\columnwidth]{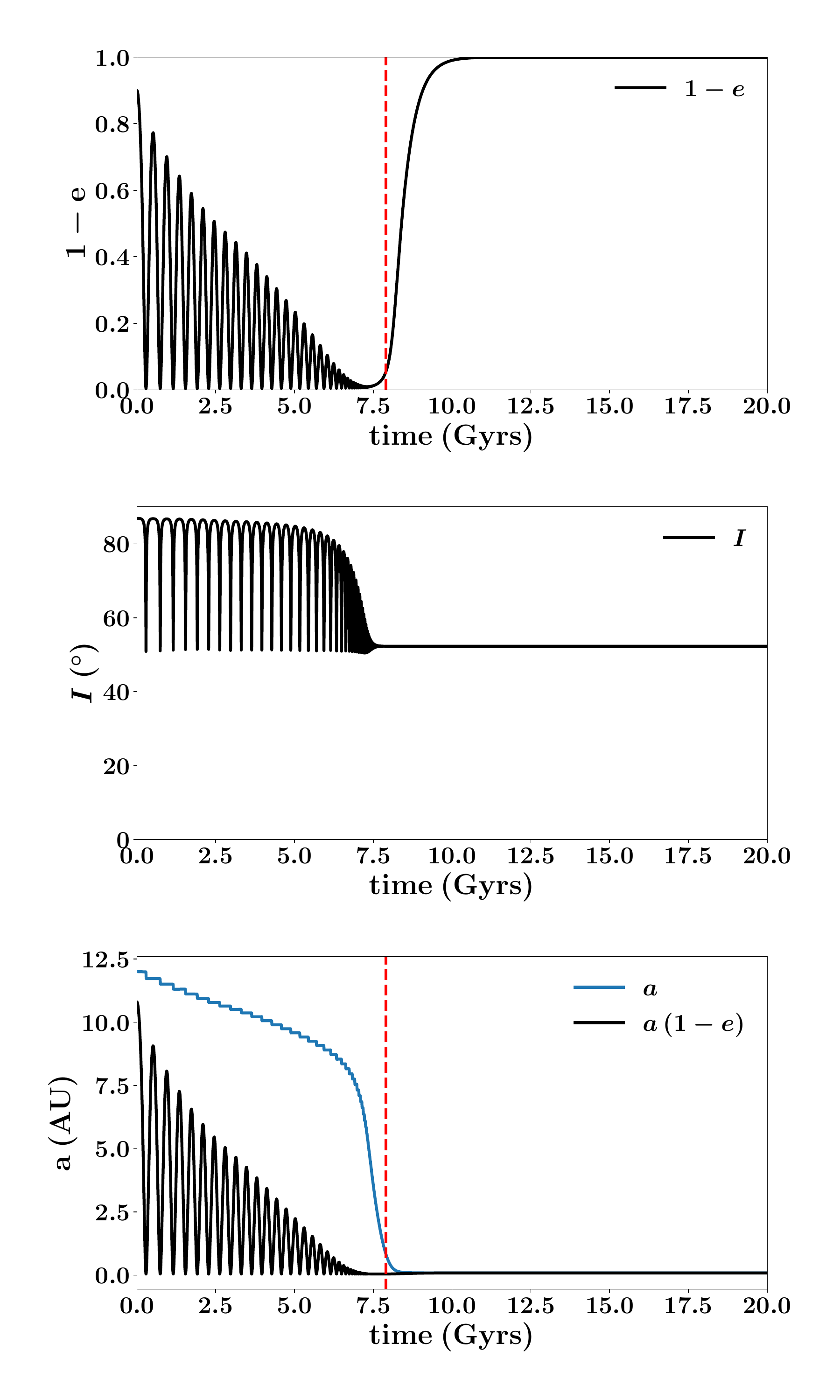}
    \caption{A plausible dynamical history of Gaia~3b according to high-eccentricity migration induced by ZLK oscillations. From top to bottom, the panels display the evolution of $1-e$, mutual inclination $I$, and $a$ (blue line) and periapse distance $a(1-e)$ (black line). The simulation demonstrates a proof-of-concept that the three-body hierarchical system can reach its present-day observed values at the expected age of approximately 7.9 Gyr, marked with vertical dashed red lines in the panels.}
    \label{fig: simulation plot}
\end{figure}

\begin{equation}
\varepsilon_{\mathrm{GR}} = \frac{3G(M_{\star} + m_{\mathrm{2}})^2 a_{2}^3 (1 - e_{\mathrm{2}}^2)^{3/2}}{M_{\mathrm{3}} a^4 c^2}.
\end{equation}
Based on these equations, we conclude that the system would have reached its present-day orbital parameters via ZLK oscillations under the condition that ${e_{\mathrm{init},\max}\gtrsim0.996}$ and a corresponding $I_\mathrm{init} \gtrsim 86.9^{\circ}$ for our choice of initial semi-major axis ${a_{\mathrm{init}}} = 12$ au. 

To simulate the planet's history, we initialize the system with ${e_{\mathrm{init}}} = 0.1$ for Gaia~3b and an initial eccentricity of the binary companion ${e_{2,\mathrm{init}}} = 0.1$. All other binary companion parameters are set to those listed in Section \ref{sec:intro}. We set the planet's mass to $m_2\sin{i}$ yielded by the RV fit. Tidal parameters are tuned such that the system reaches its current eccentricity and semi-major axis at approximately 7.9 Gyrs, and we run the simulation for 20 Gyr.

Figure \ref{fig: simulation plot} displays the results of the simulations.  The planet goes through multiple episodes of maximum eccentricity, during which strong tidal interactions are present at close periastron passages. These tidal forces gradually dissipate the orbital energy of the system. The system eventually reaches the present-day close-in orbit, after the semi-major axis decreases below the critical threshold such that short-range forces dominate and ZLK oscillations are quenched. Tidal dissipation continues and causes the planet to reach complete circularization at approximately 11 Gyr. We observe that the mutual inclination between the orbit of the planet and the perturber ranges from approximately $50^\circ$ to $87^\circ$ before tidal circularization and reaches a constant value after ZLK oscillations have quenched.  The dynamical simulation demonstrates that ZLK oscillations can reproduce the present-day orbital architecture at the expected age of approximately 7.9 Gyr. 

We highlight again that these simulations offer a proof of concept of a history of high-eccentricity migration induced by ZLK oscillations, but further theoretical work is encouraged. In particular, the Gaia~3 system may be another exciting test bed for evaluating mechanisms of tidal dissipation (equilibrium tides and chaotic dynamical tides) similarly to the analysis done for the super-eccentric giant planet TIC 241249530 b \citep{Gupta2024, Liveoak2026}.


\section{Discussion and Conclusions}\label{sec:conclusions}

High-resolution spectroscopic follow-up continues to be vital for assessing the validity of astrometrically identified planets. With our newly obtained NEID spectra and RVs spanning an entire periastron passage of Gaia 3b, we refine the uncertainty on the orbital period to \resP, and affirm the super-eccentric (\resEcc) nature of the planetary companion. Our additional RVs provide a precise understanding of the orbital parameters which will help us interpret the Gaia DR4 astrometric data when it is released. 

We generate synthetic realizations of the astrometric time series with \texttt{gaiamock} \citep{gaiamock-El-Badry2024} across multiple mass ratios and constrain the plausible parameter space for the mass of the nearby companion that would have produced the RUWE reported by Gaia to the planetary or equal-mass binary scenarios. We show the equal-mass binary case is unlikely by showing that the bisector measurements of the cross-correlation functions derived from our high resolution NEID spectra are broadly inconsistent with expectations from our two gaussian model, along with the position of Gaia 3 on a Gaia color-magnitude diagram being consistent with other single spectral standard stars of the same spectral type. Through this, we provide an alternative confirmation of the planetary nature of Gaia~3b. 

The confirmation and the relative proximity of Gaia~3b ($d\approx35$ pc) leads us to speculate that the true number of super-eccentric planets orbiting nearby Sun-like stars is greater than the current observed population. Further detections, like with the Gaia DR4 astrometric time-series paired with spectroscopic observations, could help us better constrain the eccentricity distribution for planets with $e > 0.9$ and begin to discern between the various dynamical pathways that shape the population of hot Jupiters. Given the presence of the distant stellar companion, we highlight that  ZLK oscillations provide a plausible pathway for the excitation of Gaia~3b's eccentricity and a future as a hot Jupiter. With the upcoming Gaia DR4 astrometry and our RVs we may precisely determine the mass of the planet and ascertain if Gaia 3b adheres to the trend in \citet{Gupta2024} of the observed population of high-eccentricity planets having preferentially higher masses (>2 M$_{\mathrm{Jup}}$). Our improved precision on the orbital period of Gaia~3b's orbit allows for better predictions of future periastron passages which enhances the capability to search for radio emission induced by possible star-planet interactions with LOFAR 2.0 \citep{Callingham2024}. An enhanced knowledge of the super-eccentric tail of the exoplanet eccentricity distribution offers exciting avenues for exploring the orbital dynamics that shape planetary architectures and lead to the formation of hot Jupiters. 

\begin{deluxetable*}{lcccc}[h]
\tabletypesize{\scriptsize}
\tablecaption{Orbital Parameters for Known Super-Eccentric Giant Planets}
\label{tab:lit_planets}
\tablehead{
\colhead{Name (Reference)} & \colhead{$P$ [days]} & \colhead{$e$} & \colhead{$d$ [pc]} & \colhead{$m_p$ or $^{*}$$m_p\sin i$ [M$_{\mathrm{Jup}}$]} } 
\startdata
$^{\dagger}$HD 20782b (1) & $ 597.064 \pm 0.026 $ & $0.9500 \pm 0.0010$ & $35.986^{+0.041}_{-0.041}$ & $^{*}$$1.49^{+0.11}_{-0.11}$ \\ 
$^{\dagger}$HD 28254b (2) & $ 1333.0 \pm 4.0 $ & $ 0.950^{+0.030}_{-0.040} $ & $55.288^{+0.094}_{-0.093}$ & $3.8^{+3.0}_{-2.2}$   \\ 
$^{\dagger}$Gaia~3b (This Work) & $331.526 \pm 0.015$ & $0.9462 \pm 0.0017$ & $35.399^{+0.015}_{-0.015}$ & $^{*}$$0.740 \pm 0.024$   \\
$^{\dagger}$TIC 241249530b (3) & $ 165.77190^{+0.00027}_{-0.00028} $ & $ 0.94120 \pm 0.00090 $ & $324.5^{+3.7}_{-3.6}$ & $4.98^{+0.16}_{-0.18}$  \\
$^{\dagger}$HD 80606b (4) & $111.43670 \pm 0.00040$ & $0.93226^{+0.00064}_{-0.00069}$ & $66.47^{+0.16}_{-0.16}$ & $4.12^{+0.10}_{-0.10}$  \\
HD 80869b (5) & $1752.9^{+37.0}_{-33.6}$ & $0.924^{+0.053}_{-0.094}$ & $85.40^{+0.54}_{-0.53}$ & $5.07^{+2.55}_{-0.57}$  \\
Kepler-1704b (6) & $988.88112 \pm 0.00091$ & $0.9206^{+0.010}_{-0.016}$ & $835.5^{+9.3}_{-9.1}$ & $4.16^{+0.29}_{-0.28}$  \\
\enddata
\tablenotetext{}{\textbf{Note.} Columns are orbital period ($P$), eccentricity ($e$), distance ($d$), and the planet's mass ($m_p$) or minimum mass ($m_p\sin i$, designated by a $^*$). $^{\dagger}$This star has a stellar companion. References: (1) \citet{Udry2019}, (2) \citet{Philipot2023}, (3) \citet{Gupta2024}, (4) \citet{Bonomo2017}, (5) \citet{Feng2022}, (6) \citet{Dalba2021}}
\end{deluxetable*}

\begin{acknowledgments}
AI was used for code development to perform analysis done in this work through inline code completion using the GPT-4.1 Copilot model from GitHub Copilot integrated directly into the Visual Studio Code IDE.

This paper contains data taken with the NEID instrument, which was funded by the NASA-NSF Exoplanet Observational Research (NN-EXPLORE) partnership and built by Pennsylvania State University. 
NEID is installed on the WIYN telescope, which is operated by the NSF's National Optical-Infrared Astronomy Research Laboratory, and the
NEID archive is operated by the NASA Exoplanet Science Institute at the California Institute of Technology. NEID is funded by NASA through JPL contract 1547612 and the NEID Data Reduction Pipeline is funded through JPL contract 1644767.
NN-EXPLORE is managed by the Jet Propulsion Laboratory, California Institute of Technology under contract with the National Aeronautics and Space Administration. We thank the NEID Queue Observers and WIYN Observing Associates for their skillful execution of our observations. E.F., S.M, M.D., and P.R. acknowledge support from NASA through the Exoplanet (XRP) research program grant 80NSSC24K0155.
S.P.H acknowledges that part of this work was performed at the Jet Propulsion Laboratory, California Institute of Technology, sponsored by the United States Government under the Prime Contract 80NM0018D0004 between Caltech and NASA. S.H.A. and K. K. K. acknowledge support from the Independent Research Fund Denmark (DFF) through grant 10.46540/5241-00072B.

Computations for this research were performed on the Pennsylvania State University's Institute for Computational and Data Sciences Advanced Cyberinfrastructure (ICDS-ACI). This content is solely the responsibility of the authors and does not necessarily represent the views of the Institute for Computational and Data Sciences. The Center for Exoplanets and Habitable Worlds and the Penn State Extraterrestrial Intelligence Center are supported by the Pennsylvania State University and the Eberly College of Science. This work has made use of data from the European Space Agency (ESA) mission Gaia, processed by the Gaia Data Processing and Analysis Consortium (DPAC). Funding for the DPAC has been provided by national institutions, in particular the institutions participating in the Gaia Multilateral Agreement. This research has made use of the SIMBAD database, operated at CDS, Strasbourg, France, and NASA's Astrophysics Data System Bibliographic Services.

Based in part on observations at Kitt Peak National Observatory, NSF’s NOIRLab, managed by the Association of Universities for Research in Astronomy (AURA) under a cooperative agreement with the National Science Foundation. Kitt Peak National Observatory sits atop I’oligam Du’ag (Manzanita Shrub Mountain). Astronomers are honored to be permitted to conduct scientific research on the sacred mountain located in the homelands of the Schuk Toak District within the Tohono O’odham Nation. We honor their past, present, and future generations, who have lived here for time immemorial and will forever call this place home.
We also express our deepest gratitude to Zade Arnold, Joe Davis, Michelle Edwards, John Ehret, Tina Juan, Brian Pisarek, Aaron Rowe, Fred Wortman, the Eastern Area Incident Management Team, and all of the firefighters and air support crew who fought the recent Contreras fire. Against great odds, you saved Kitt Peak National Observatory.

The Pennsylvania State University campuses are located on the original homelands of the Erie, Haudenosaunee (Seneca, Cayuga, Onondaga, Oneida, Mohawk, and Tuscarora), Lenape (Delaware Nation, Delaware Tribe, Stockbridge-Munsee), Shawnee (Absentee, Eastern, and Oklahoma), Susquehannock, and Wahzhazhe (Osage) Nations.  As a land grant institution, we acknowledge and honor the traditional caretakers of these lands and strive to understand and model their responsible stewardship. We also acknowledge the longer history of these lands and our place in that history.

\end{acknowledgments}

%

\vspace{5mm}
\facilities{: Gaia, WIYN (NEID), NASA Exoplanet Archive \citep{NEA-CompositePlanetTable}}


\software{\texttt{astropy} \citep{AstropyCollaboration2013,AstropyCollaboration2018,AstropyCollaboration2022}, \texttt{barycorrpy} \citep{kanodia2018,WrightandEastman2014}, \texttt{matplotlib} \citep{Hunter2007}, \texttt{numpy} \citep{Harris2020}, \texttt{radvel} \citep{Fulton2018}, \texttt{scipy} \citep{Oliphant2007}, \texttt{gaiamock} \citep{gaiamock-El-Badry2024}, \texttt{SERVAL} \citep{zechmeister2018,stefansson2022}, \texttt{Octofitter.jl} \citep{Thompson2023_Octofitter}, \texttt{SecularMultiple} \citep{HamersPortegiesZwart2016,Hamers2018,Hamers2020}, \texttt{GORP-masses} \citep{Giovinazzi2022}, \texttt{emcee} \citep{emcee_ForemanMackey2013}     }



\bibliography{references}{}
\bibliographystyle{aasjournalv7}

\appendix
\setcounter{table}{0}
\renewcommand{\thetable}{A\arabic{table}}
\twocolumngrid

\section{Additional RV Information}\label{appendix:RVFitInfo}

\subsection{Radial Velocities}\label{appendix:RVs}
We provide our NEID radial velocities and associated uncertainties measured using a modified version of the template-matching algorithm \texttt{SERVAL} in Table \ref{table:neid_rvs}.

\begin{deluxetable}{cccc}[h]
\tablecaption{NEID \texttt{SERVAL} RVs for Gaia~3b
\label{table:neid_rvs}}
\tabletypesize{\scriptsize}
\tablehead{\colhead{BJD [days]} &  \colhead{RV [$\text{m}\:\text{s}^{-1}$]}     & \colhead{$\sigma_{\rm{RV}}$ [$\text{m}\:\text{s}^{-1}$]} & \colhead{NEID Run \#}}
\startdata      
2460271.98777 & -29.2 & 1.4 & RUN2 \\
2460309.96027 & 6.4 & 1.1 & RUN2 \\
2460319.9964 & 13.1 & 1.1 & RUN2 \\
2460338.83927 & 7.0 & 1.5 & RUN2 \\
2460340.99301 & 5.1 & 0.9 & RUN2 \\
2460396.80676 & 2.9 & 2.3 & RUN2 \\
2460407.63452 & 1.1 & 1.7 & RUN2 \\
2460441.97114 & -1.3 & 1.0 & RUN2 \\
2460459.92558 & -5.1 & 1.2 & RUN2 \\
2460473.935 & 2.9 & 1.3 & RUN2 \\
2460489.83295 & -1.94 & 0.94 & RUN2 \\
2460499.87871 & -7.5 & 1.2 & RUN2 \\
2460556.71591 & 27.7 & 1.3 & RUN3 \\
2460557.72249 & 27.7 & 1.3 & RUN3 \\
2460567.66862 & 31.5 & 1.8 & RUN3 \\
2460608.02402 & 11.1 & 1.2 & RUN3 \\
2460615.61965 & -1.5 & 1.3 & RUN3 \\
2460623.59005 & -52.1 & 1.1 & RUN3 \\
2460625.56872 & -112.6 & 5.9 & RUN3 \\
2460625.96527 & -74.4 & 1.2 & RUN3 \\
2460626.59243 & -9.1 & 2.6 & RUN3 \\
2460627.00583 & 8.31 & 0.96 & RUN3 \\
2460628.01812 & 30.40 & 0.84 & RUN3 \\
\enddata
\tablenotetext{}{}
\end{deluxetable}

\subsection{Predicted Periastron Passage Dates}
Using our updated orbital parameters we compute and provide the next 10 dates for the expected timing of periastron passage  in Table \ref{tab:peri_passages}. 
\begin{deluxetable}{lc}[h]        
\tablecaption{Next 10 predicted periastron passage times for Gaia~3b\label{tab:peri_passages}}
\tabletypesize{\scriptsize}       
\setlength{\tabcolsep}{3pt}       
\tablewidth{0pt}                  

\tablehead{
  \colhead{Julian Day} & \colhead{UTC Date-Time (Mean)}  
}
\startdata
$2461288.69 \pm 0.05$  & September 5, 2026 - 04:32:56  \\ 
$2461620.22 \pm 0.06$  & August 2, 2027 - 17:10:07  \\ 
$2461951.74 \pm 0.08$  & June 29, 2028 - 05:47:19  \\ 
$2462283.27 \pm 0.09$  & May 26, 2029 - 18:24:31 \\ 
$2462614.79 \pm 0.10$ & April 23, 2030 - 06:57:36 \\
$2462946.32 \pm 0.12$ & March 20, 2031 - 19:40:48 \\
$2463277.85 \pm 0.13$ & February 15, 2032 - 08:24:00 \\
$2463609.37 \pm 0.15$ & January 11, 2033 - 20:52:48 \\
$2463940.90 \pm 0.16$ & December 9, 2033 - 09:36:00 \\
$2464272.42 \pm 0.18$ & November 5, 2034 - 22:04:48\\
\enddata
\end{deluxetable}

\end{document}